\documentclass[a4paper,fleqn,usenatbib]{mnras}
\usepackage[T1]{fontenc}
\usepackage{ae,aecompl}
\usepackage{graphicx}
\usepackage{amsmath}
\usepackage{amssymb}
\usepackage{here}
\usepackage{color}
\title[Radio broadband visualization of simulated spiral galaxies I]{Radio broadband visualization of global three-dimensional magneto-hydrodynamical simulations of spiral galaxies II. Faraday Depolarization from 100~MHz to 10~GHz
}
\author[M. Machida et al.]{
M. Machida,$^{1}$ 
T. Akahori,$^{2,3}$ 
K. E. Nakamura$^{4}$,  
H. Nakanishi$^{2,5,6}$  
and M. Haverkorn $^{7}$ 
\\
$^{1}$Department of Physics, Faculty of Sciences, Kyushu University,  744 Motooka, Nishi-ku, Fukuoka, 819-0395, Japan 
\thanks{mami@phys.kyushu-u.ac.jp} \\
$^{2}$Graduate School of Science and Engineering, Kagoshima University, Korimoto 1-21-35, Kagoshima 890-0065, Japan\\
$^{3}$National Astronomical Observatory of Japan, 2-21-1 Osawa, Mitaka, Tokyo 181-8588, Japan \\
$^{4}$Kyushu Sangyo University, 3-1 Matsukadai 2-chome, Higashi-ku, Fukuoka, 813-8503, Japan\\
$^{5}$Institute of Space and Astronautical Science, Japan Aerospace Exploring Agency, 3-1-1 Yoshinodai, Sagamihara, Kanagawa 252-5210, Japan\\
$^{6}$SKA Organization, Jodrell Bank Observatory, Lower Withington, Macclesfield, Cheshire SK11 9DL, UK\\
$^{7}$1Department of Astrophysics / IMAPP, Radboud University Nijmegen, PO Box 9010, 6500 GL Nijmegen, The Netherlands \\
}

\date{Accepted XXX. Received YYY; in original form ZZZ}
\pubyear{2016}
\begin{document}
\label{firstpage}
\pagerange{\pageref{firstpage}--\pageref{lastpage}}
\maketitle

\begin{abstract}

Observational study of galactic magnetic fields is limited by projected observables. Comparison with numerical simulations is helpful to understand the real structures, and observational visualization of numerical data is an important task. \cite{mac2017} have reported Faraday depth maps obtained from numerical simulations. 
They showed that the relation between azimuthal angle and Faraday depth depends on the inclination angle. 
In this paper, we investigate 100\,MHz to 10\,GHz radio synchrotron emission from spiral galaxies, using the data of global three-dimensional magneto-hydrodynamic simulations. We model internal and external Faraday depolarization at small scales and assume a frequency independent depolarization. 
It is found that the internal and external Faraday 
depolarization becomes comparable inside the disk and the dispersion of Faraday depth becomes about $4\,{\rm rad\,m^{-2}}$ for face-on view and $40\,{\rm rad\,m^{-2}}$ for edge-on view, respectively. The internal depolarization becomes  ineffective in the halo. Because of the magnetic turbulence inside the disk, frequency independent depolarization works well and the polarization degree becomes 0.3 at high frequency. When the observed frequency is in the 100 MHz band, polarized intensity vanishes in the disk, while that from the halo can be observed. Because the remaining component of polarized intensity is weak in the halo and the polarization degree is about a few \%,  it may be difficult to observe that component. 
These results indicate that the structures of global magnetic fields in spiral galaxies could be elucidated, if broadband polarimetry such as that with the Square Kilometre Array is achieved.
\end{abstract}


\section{Introduction}
It is known that spiral galaxies have magnetic fields with strengths of a few $\mu {\rm G}$ on average \cite[e.g.,][]{soi2011}, and the energy of the global, ordered magnetic field is comparable to the energy of the local, turbulent magnetic field. 
The global fields topologies of spiral galaxies have been studied using by the relation between azimuthal direction and the Rotation Measure (RM), such as Axisymmetric Spiral and Bisymmetric Spiral \citep{fuj1987, chi92, kra1990}.    
The high spatial resolutions obtained with radio interferometric observations, such as with the Jansky Very Large Array (JVLA), allows observations of smaller scale components and investigation of the detail structures in RM \cite[e.g.,][and references therein]{bec2016}.
%
For example, the magnetic fields of M51 which has a clear grand-design spiral show a radial distribution 25--10$\mu\,{\rm G}$ from center to the outer edge \citep{fle2011}.  
The structure of the halo magnetic fields are reported in face-ons/edge-ons  \cite[e.g.,][]{kra2014, mul2017, mul2018}.  
Maintaining such magnetic field strength in a normal spiral galaxy for billions of years requires a maintenance mechanism that acts against magnetic dissipation.  As a model of the maintenance mechanism, \citet{par1970} and \citet{par1971} proposed a galactic dynamo that converts the kinetic energies of the galactic differential rotation and associated turbulent motion into magnetic energy.  The galactic dynamo can amplify the initial weak seed magnetic field and maintains microgauss magnetic fields over $10^8$ years.  
Extended models based on galactic dynamo theory were proposed in the 1980s.  
One of the key constraints in the mean-field dynamo theory is the conservation of magnetic helicity \cite[e.g.,][and references therein]{bla2001}. 
The growth of magnetic helicity of opposite signs between the large-scale and small scale magnetic fields suppress the dynamo action.  
The small scale magnetic helicity fluxes should be removed to avoide the suppression and there are some candidate mechanisms such as 
advection of magnetic fields by outflow, turbulence produced by differential rotation and diffusion. 
In order to consider the effect of these process, recent theory of the mean field dynamo include the connection between disk and halo \citep{shu2006, pra2016} which is 
considered to bethe effect of the supernova explosions and magneto-rotational instability (MRI) \citep{bal1991} They concluded that the growth rate of the turbulence by supernova explosions is faster than that of MRI. The explosions of the magnetic helicity make a corona which has a large scale magnetic field. 
The magnetic field strength in the corona is much weaker than that in the disk.

However, these works do not consider the feedback of the magnetic fields and 
the influence of magnetic fields on the motion of the gas cannot be ignored \citep{bal1991} in the diferentially rotating system.  
To build a galaxy model that takes into account the effects of magnetic fields on the gas motion, several global simulations of a galactic gas disk have been carried out (\cite{nis2006, han2009, mac2009, mac2013}).  \citet{mac2013} performed a three-dimensional (3D) magneto-hydrodynamic (MHD) simulations of the galactic gas disk, and found that the evolution of the galactic disk can be described by the dynamo effect according to the MRI and Parker instability.
 As the MRI grows with the magnetic pressure reaching 10\,\% of the gas pressure, 
the azimuthal magnetic field 
is amplified.  Then, where magnetic pressure reaches 20\,\% of the gas pressure, the magnetic flux gradually floats out from the disk due to the Parker instability.  
Since this flux lifting causes the dynamo action, the disk magnetic field is again amplified.  
Even if only weak magnetic field is assumed at the beginning, numerical results show global spiral magnetic arms with turbulence.  
The direction of the azimuthal magnetic field inverts quasi-periodically in both radial and vertical directions. 
The magnetic fields amplified inside the disk are supplied to the halo constantly. 
This mechanism has, however, not yet been proven by observations; it is still difficult to measure global magnetic fields in galaxies even with current large observational facilities such as the Jansky Very Large Array and LOFAR.  
Wideband and high--spatial--resolution observations with the Square Killometre Array (SKA) will reveal complex structures that reflect the MRI and Parker instability.  Therefore, theoretical predictions prior to observations are important.

\citet{mac2017} (hereafter Paper I) obtained Faraday depth (FD) maps and synchrotron intensity at high frequency using the results of \citet{mac2013}. 
It is found that the appearance of global magnetic field depends on the viewing angle; the magnetic fields appears as a hybrid type which consist of an axisymmetric spiral and higher modes for a face-on view,  partially a ring-like structure at an inclination of $\sim 70 \degr$, and a structure parallel to the disk for the edge-on view. The magnetic vector seen at centimeter wavelengths traces the global magnetic field inside the disk. 

In this paper, we investigate 100\,MHz to 10\,GHz radio synchrotron emission from spiral galaxies to provide a theoretical model of global magnetic fields in spiral galaxies.  We aim to clarify the relation between projected observables and the actual 3D structure of galactic magnetic fields using our sophisticated model. We would also like to clarify the effect of Faraday depolarization.  We introduce the calculation method in Section 2. The numerical results are shown in Section 3, followed by discussion and summary in Sections 4 and 5, respectively.

\section{Observational Visualization}

Global MHD simulation of a galactic gaseous disk by \citet{mac2013} are adopted to our model galaxy, similar to Paper I. 
We adopted the cylindrical coordinate system $(r, \phi, z' )$. The simulated region is $r\,{\rm kpc} < 56 $ in the radial direction, $|z'|\,{\rm kpc} < 10$ in the vertical direction and $0 \leq \phi \leq 2 \pi$ in the azimuthal direction. The spatial resolution where we focus on this paper is $\Delta r=50$\,pc, $\Delta z'=10$\,pc, and $\Delta \phi = 2\pi/256$, respectively. 
The units of the numerical calculation such as the length, velocity and density are $r_0 = 1{\rm kmp}$, 
$v_0 = 207~{\rm km/s}$ and $\rho_0 = 1~{\rm cm^{-3}}$, respectively. 
The unit of magnetic field strength is defined by the plasma $\beta$ as $B_0 = \sqrt{\rho_0 v_0^2}= 26 \mu {\rm G}$. 
After the magnetic turbulent was amplified by MRI, the disk reached the quasi-steady state. At that time, the averaged density becomes $0.01~{\rm cm^{-3}}$ and magnetic field strength is about 5 $\mu {\rm G}$. The gas temperature inside the disk is about a few $10^5 {\rm K}$ because the gas of the disk assumed the adiabatic. 
The plasma $\beta$ ranges from about a few times $10^{-1}$ to $10^{3}$. 
Only less than 1\% voluume of the galactic disk has the low $\beta$ ($\beta <5$). 
Otherwise, the plasma $\beta$ is typically 10--100 in the disk.
The details of our numerical simulation are shown in \citet{mac2013} and appendix A in Paper I. 

The model galaxy is assumed an external galaxy and we calculate the Faraday rotation measure and the Stokes parameter of the radio band. 
Toward future radio polarimetry, we consider broadband observations between 100\,MHz to 10\,GHz throughout the paper.  Hereafter, we refer to the Faraday depth (FD) as a line-of-sight (LOS) integral of the LOS component of magnetic field weighted with the density, while RM is given by wavelength dependence of the polarization angle (PA).

The visualization of the simulation is a cube with size 20\,kpc and coordinates $(x,y,z)$ centered on the center of the galaxy. We construct two-dimensional maps of $200\times200$\,pixels centered on the projected center of the galaxy, which results in an image resolution of 100\,pc, corresponding to $\sim2"$ for a galaxy in the Virgo cluster at a distance of $\sim 20$\,Mpc.

\subsection{Model of Depolarization}

In the visualization, the spatial resolution of numerical data is a main factor of uncertainty. We cannot directly reproduce depolarization features \citep{Burn66, sok98, ars2011, shn2014} caused by structures below the resolution. Here, the frequency independent depolarization determines the intrinsic polarization degree (PD), which can be parameterized using strengths of the mean and turbulent magnetic fields. The differential Faraday rotation depolarization primarily depends on mean LOS magnetic fields, whose effect can be naturally considered in numerical integration along the LOS, while an effect from unresolved turbulent LOS magnetic fields should be on a higher order.  As for the beam depolarization, both internal and external Faraday dispersion depolarization caused by unresolved turbulent magnetic fields can be significant and we therefore describe it adequately.

Unless the resolution is much higher than the characteristic scale of turbulent magnetic fields, polarization will be underestimated because of significant Faraday dispersion depolarization on sub-resolution scales. Reproducing turbulence on a parsec scale, which is expected to be the characteristic scale in spiral arms (e.g., \citealt{hav2006, bec2007}), is however expensive for our simulation of a whole galaxy with a scale of a few tens of kpc. Therefore, visualization of global MHD simulation of spiral galaxies is essentially quite challenging, particularly at low frequencies.

In our numerical data, the azimuthal magnetic field power spectrum obeys the Kolmogorov law with a power-law index $a\sim -5/3$ for wavenumbers $n>6$. The spectra of the other field components and the gas density near the equatorial plane also indicate power-law spectra with an index $-5/3 \lesssim a \lesssim -1$ for $n>10$. The simulation thus solved the forcing range and part of the inertial range of turbulence. 
Since the amplitude of the power spectrum is still enough values on smaller scale, the turbulence on smaller scale than the spatial resolution is considered assuming the Kolmogrov law, but not neglected. 
On the other hands, the power-law index for the halo components becomes smaller than that in the disk and significant peaks do not appear. 
Therefore, we adopt the laws of Faraday dispersion depolarization, and replace the dispersion of FD, $\sigma_{\rm FD}$, for a certain image pixel into fluctuation of FDs around the pixel. This model makes an uncertainty on this work, which however is within a factor of about 2 (see Section 4.1).

\subsection{Procedure of Visualization}
\
The procedure of visualization is similar to Paper I except for the treatment of Faraday depolarization. 
The integration for the image pixel at $(i, j)$ is performed as follows. The local FD within a certain computational cell is given by $FD_{i,j,l} =0.81\,n_{\rm e} B_{\parallel} \Delta l$, where $\Delta l=100$\,pc is the line element. The electron density, $n_{\rm e}$, and the LOS component of magnetic field, $B_{\parallel}$, are derived from interpolation of the nearest eight cells of the cylindrical coordinate in the data. The FD along the LOS is calculated by integrating local FDs. Here, we separately integrate FDs of the global mean field along the LOS $\overline{B}_{\parallel}$ and the local turbulent field $b_{\parallel} = B_{\parallel} - \overline{B}_{\parallel} $ as
\begin{equation}
FD_{{\rm ave},i,j,k} =\sum_{l=1}^{k} 0.81\, n_{\rm e}  \overline{B}_{\parallel} \,\Delta l,
\end{equation}
and
\begin{equation}
FD_{{\rm turb},i,j,k} =\sum_{l=1}^{k} 0.81\, n_{\rm e}  \,b_{\|} \,\Delta l,
\end{equation}
respectively. The mean field at each cell is computed by averaging magnetic fields inside the $\pm N$ cells in each directions. We adopt $N=5$, because the resultant length is a few times larger than the size of image pixel $\sim 100\,{\rm pc}$. Based on the model described in Section 3.1, the standard deviation of FD in front of $k$-th grid in $(i, j)$ pixel, $\sigma_{{\rm FD},i,j,k}$, is evaluated by fluctuation of FDs up to $k$-th grids in the eight nearest cells of the image pixels  ($m$) surrounding $(i,j)$ pixel as
\begin{equation}
\sigma_{{\rm FD},i,j,k}^2 =\frac{1}{8} \sum_{m=1}^8 (FD_{{\rm turb},i,j,k,m} - FD_{{\rm ave},i,j,k,m})^2.
\end{equation}
We also integrate the optical depth,
\begin{equation}
\tau_{i,j,k} = 8.235\times 10^{-2} \nu^{-2.1} \sum_{1}^{k} T^{-1.35} n_{\rm e}^2 \Delta l,
\end{equation}
so as to consider free--free absorption. Here $T$ is the electron temperature derived from interpolation of the nearest eight cells of the image pixels in the data, and $\nu$ is the frequency.

The internal and external Faraday dispersion depolarization are estimated from \citep{ars2011}:
\begin{equation}
D_{\rm i} = \left( \frac{1 - \exp{(- 2 \sigma_{FD_{i,j,\delta l}}^2 \lambda^4)}}
            {2 \sigma_{FD_{i,j,\delta l}}^2 \lambda^4} \right), 
\end{equation}
and
\begin{equation}
D_{\rm e} = \exp{( - 2 \sigma_{\rm{FD}, i,j,k}^2 \lambda^4)},
\end{equation}
respectively, where $\lambda$ is the wavelength and $\sigma_{FD_{i,j,\delta l}}$ is the dispersion of the local $FD_{i,j,\delta l}$.

Specific Stokes parameters are given by the formula in the literature \citep{sun08, wae09}. With the cosmic-ray electron energy spectral index, $p$, and the coefficients, $g_1(p)$ and $g_2(p)$ (see Paper I, Appendix B), observable specific Stokes parameters of synchrotron radiation emitted at $k$-th grid are given by
\begin{equation}
\Delta I= g_1(p) F dl,
\end{equation}
\begin{equation}\label{eq:dQ}
\Delta Q=g_2(p) F \cos{(2\chi_{i,j,k})} \, W D_{\rm i} D_{\rm e} \Delta l,
\end{equation}
and
\begin{equation}\label{eq:dU}
\Delta U=g_2(p) F \sin{(2\chi_{i,j,k})} \, W D_{\rm i} D_{\rm e} \Delta l,
\end{equation}
where
\begin{equation}
F = C(r) B_{\bot}^{(1-p)/2} (2 \pi\nu)^{(1+p)/2} e^{-\tau_{i,j,k}}.
\end{equation}
The term $e^{-\tau_{i,j,k}}$ and $W$ give free--free absorption of synchrotron emission and frequency independent depolarization, respectively.  

Faraday rotation of the PA in Eqs. (\ref{eq:dQ}) and (\ref{eq:dU}) is given by
\begin{equation}
\chi_{i,k} = \frac{1}{2}\cos^{-1}{\left( \frac{B_x}{|B_{\bot}| } \right) } + FD_{{\rm ave},i,j,k} \lambda^2 + FD_{{\rm turb},i,j,k} \lambda^2,
\end{equation}
where the first term gives the intrinsic PA. We integrate $\Delta I$, $\Delta Q$, and $\Delta U$ along the LOS to obtain the specific Stokes parameters ($I$, $Q$, and $U$). 
The unit of Stokes parameter is $\mu {\rm Jy/beam}$ where we assume a $10$ arcsec beam size in diameter.

We also present the PA,
\begin{equation}\label{eq:chi}
\chi = \frac{1}{2} \tan^{-1} \left(\frac{U}{Q} \right), 
\end{equation}
and the PD,
\begin{equation}
{\rm PD} = \frac{P}{I}=\frac{\sqrt{Q^2+U^2}}{I},
\end{equation}
where we do not consider circular polarization so that it is weak in the context of this study.

We assume that the cosmic-ray has equilibrium energy density with the magnetic energy. 
In order to obtain the number density of the cosmic-ray, we have assumed the range of the Lorentz factor for cosmic-ray electrons 
is 200--3000 \citep{aka2018}.  
The energy spectral index $p=3$ is adopted as a typical value \citep{sun08}.

\section{Results}

Because the result of the simulation data we assumed is similar data set of Paper I, the results at 8\,GHz are same as that of Paper I.  
Below, we first determine the fiducial unit density of the MHD simulation data, then show the results of face-on view and edge-on view.

\subsection{Determination of the fiducial unit density}

\begin{figure}
\begin{center}
\includegraphics[width=8cm]{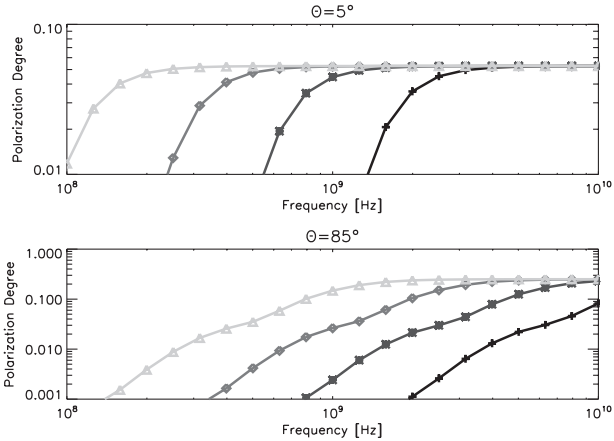} 
\end{center}
\caption{
PDs of the emission from an arm region for the inclination angle of $5 \degr$ (nealy face-on, top panel) and from a disk region for the inclination angle of $85 \degr$ (nearly edge-on, bottom panel), as a function of the frequency. The symbols denote the cases with different unit densities; $3\times 10^{-3}\,\rm{cm}^{-3}$ (triangle), $0.01\,\rm{cm}^{-3}$ (diamond), $0.03\,\rm{cm}^{-3}$ (cross), and $0.1\,\rm{cm}^{-3}$ (plus).
}\label{fig1}
\end{figure}

In our MHD simulation data, because the density is a free parameter, we can assume its value freely. 
And the magnetic field strength and the synchrotron intensity strongly depend on the unit density of the data. 
Therefore, we check the dependence of the unit density for the observables and determine the fiducial unit density we assumed.  
Fig.\,\ref{fig1} shows the dependence of the PD on the unit density. As representative cases, we chose an arm region for an inclination angle of $5 \degr$ at (1.5\,kpc, 0\,kpc) and a disk region for an inclination angle of $85 \degr$ 
at (3\,kpc, 0\,kpc). The PDs in this panel is measured on one pixel. 
The symbols denote the cases with different unit densities; they increase from left to right.

According to the previous radio observations introduced in Section 1, PDs of less than 5\% were observed at frequencies lower than 1\,GHz in face-on galaxies, and no polarization was detected at 151\,MHz in M51 \citep{fle2011, mul2014}. 
In edge-on galaxies, the typical degree of polarization at 1~GHz is a few \% in the disk and increasese to ~10--20\% in the halo. \cite[see e.g.][]{hum1991}.At 146~MHz, no polarization was detected in NGC891 with LOFAR \citep{mul2018}. 
The case with $0.03\,\rm{cm}^{-3}$ fits with the observations, therefore, we choose $0.03\,\rm{cm}^{-3}$ as the fiducial unit density in this paper. Indeed, with the unit density of $0.03\,\rm{cm}^{-3}$, the average thermal electron density in the disk is $\sim 0.03\,{\rm cm^{-3}}$   and that in the halo is $\sim 3 \times 10^{-4}\,{\rm cm^{-3}}$ which are consistent with X-ray observations of a hot disk \citep{sak2014}.

\begin{figure}
\begin{center}
\includegraphics[width=6cm]{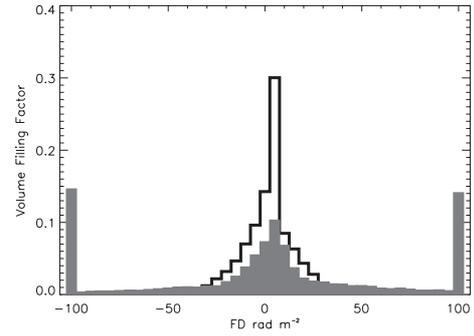} 
\end{center}
\caption{
The volume filling factor of FD. The black and gray histogram show the results in the case of $5\degr$ and $85\degr$, respectively. The horizontal axis shows FD. 
}\label{fig2}
\end{figure}

The FD contour maps (Fig. 2a and Fig. 3a in Paper I) have an upper and lower limit and the range of FD value is wide. 
The low FD region is difficult to find out, because the contrast of the FD around $0\,{\rm rad \, m^{-2}}$ is too weak to identify. 
Therefore, we show the histogram of the FDs.  
Fig.\,\ref{fig2} shows the volume filling factor of FD. The black and gray histogram show the results for inclinations of $5\degr$ and $85\degr$, respectively.  The horizontal axis show FD. The most volume zone of FD for the case of $5\degr$ is 0.3 from 5\,${\rm rad \, m^{-2}}$ to 10\,${\rm rad \, m^{-2}}$.  Although the one of the volume zone for model $85\degr$ is from 5\,${\rm rad \, m^{-2}}$ to 10\, ${\rm rad \, m^{-2}}$, |FD| > 100\,${\rm rad \, m^{-2}}$ also have 0.15. These region whose  absolute value of FD becomes greater than 100\,${\rm rad \, m^{-2}}$ corresponds to the halo and |FD|<20 ${\rm rad \, m^{-2}}$ region is the results from the disk.  Corresponding average magnetic-field strength both of disk and halo  is about 10\,$\mu{\rm G}$.

\subsection{Face-on View}

\begin{figure*}
\begin{center}
 \includegraphics[width=15cm]{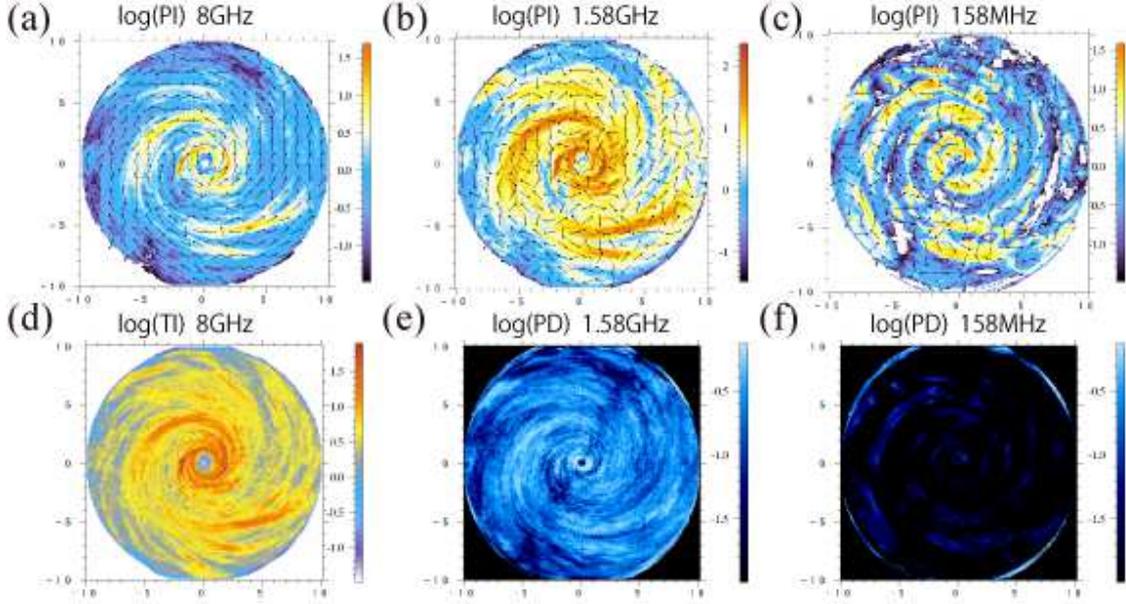}
\end{center}
\caption{
 (a), (b), (c) Simulated polarized intensity for face-on view at the observed frequencies of 8\,GHz, 1.58\,GHz and 158\,MHz, 
respectively, all in units of $\mu$Jy\,/beam. 
Arrows indicate the magnetic vector $(2\chi+90\degr)$  which 
has constant length and hence only denotes the direction of the magnetic fields. 
(d) Simulated synchrotron total intensity at the observed frequency of 8\,GHz.  
(e), (f) Polarization degree at the observed frequency of 1.58\,GHz and 158\,MHz, respectively. 
}
\label{fig3}
\end{figure*}

Polarized intensity (PI) plots are shown in Fig.\,\ref{fig3}a, \ref{fig3}b and \ref{fig3}c at the observed frequency of 8\,GHz, 1.58\,GHz and 158\,MHz, respectively.  Fig.\,\ref{fig3}d shows total synchrotron intensity at the observed frequency of 8\,GHz.  
Arrows show the magnetic vector, $2\chi+90 \degr$, with $\chi$ obtained from the Stokes Q and U (Eq.\ref{eq:chi}), 
whose direction only have a meaning.  
Polarization degrees at frequencies of 1.58\,GHz and 158\,MHz are shown in Fig.\,\ref{fig3}e and \ref{fig3}f, respectively.  As we have already shown in Paper I, total synchrotron intensity (TI) and PI(8\,GHz) reflect the spiral magnetic-field structure inside the gas disk. 
The result of the model that mean fields are stronger in the arms than in the interarm regions 
is different from observations in grand-design galaxies having strong density waves \citep{fle2011}, but such galaxies are beyond the scope of the MRI model, as mentioned in Paper I.

However, the spiral pattern becomes different at lower frequencies. That is, anti-correlation between TI and PI is found in PI(158\,MHz); TI is stronger but PI is weaker at ($1.5$\,kpc, $0$\,kpc), and TI is weaker but PI is stronger at ($2$\,kpc, $6$\,kpc). 
The polarization degree  (PD = PI/TI) distribution (Fig.\,\ref{fig3}e) is not uniform; PD becomes small where TI is relatively large and PD gradually increases at the edges of regions of strong TI. The depolarization becomes more important near the galactic center owing to the higher density and stronger magnetic field. The PD inside the magnetic spiral arms become under 1\%.  
According of these effects, the high PI regions at 158\,MHz trace the spiral magnetic field in the halo. 
As for the PI at 1.58\,GHz, the spiral pattern becomes close to that of TI, but the polarized vector rotates significantly, particularly around the disk spiral arms at which the FD is about $20$--$40$\,rad \, m$^{-2}$. These features can be ascribed to depolarization and Faraday rotation.
The muximum value of the PI at 158\,MHz is similar to that at 8\,GHz, although the positions of PI peak are different between them.  
The PI at the magnetic spiral arms becomes highest value at the high frequency, because the intensity becomes highest and Faraday depolarization does not work at such a high frequency region. On the other hands, when the obserbed frequency is 100\,MHz bands, the peak position of the PI move to the surrounding region of the magnetic spiral arm.  It is because Faraday depolarization occur in the armed region and PI decreases on the arm region. The peak intensity of PI increase from 8\,GHz to 1\,GHz, and it reduces from 1\,GHz to 100\,MHz. 
Although an increase of PI from 8~GHz to 1~GHz was observed in the outer regions of some face-on galaxies, the PI in the inner disk decreased \citep{bec2007, fle2011}. This is because the amount of Faraday depolarization computed in our calculation may be underestimated; 
the density around the central region is lower than that seen in the real spiral galaxies because the galaxy model we used have adopted an absorption boundary condition inside 200pc. 
The reason we adopted the absorption boundary is that the in-falling matter heats up the central region.

\begin{figure}
\begin{center}
 \includegraphics[width=8cm]{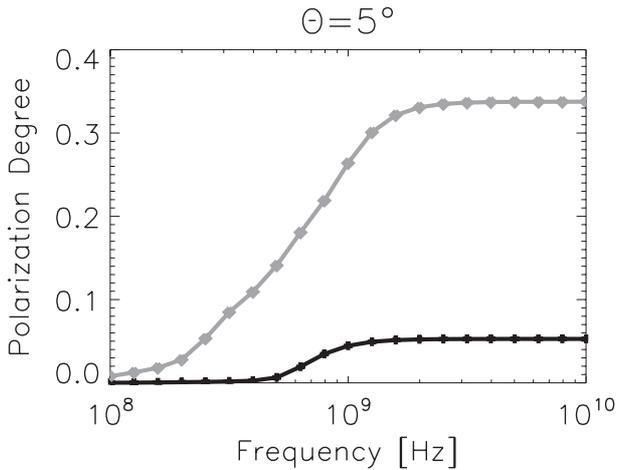}
\end{center}
\caption{
PDs of the emissions from an arm ($x$,$y$)=($1.5$\,kpc, 0.0\,kpc) (black line with crosses) and from an inter-arm ($0$\,kpc, $1.5$\,kpc) (gray line with asterisks) of $\theta=5 \degr$, as a function of the frequency.
}\label{fig4}
\end{figure}

Fig.\,\ref{fig4} shows the frequency dependence of the PD. Black and gray curves show the result in the spiral arm and in the inter-arm region. 
The PD is calculated for one pixel at ($1.5$\,kpc, $0$\,kpc) in the arm region and ($0$\,kpc, $1.5$\,kpc)  in the inter-arm region. 
Indeed, the PD decreases below a few GHz, which corresponds to FD of dozens rad \, m$^{-2}$ according to Faraday dispersion depolarization \citep{ars2011}. Below a few MHz, the PD becomes less than 0.1.
Because of the effect of frequency independent Faraday depolarization, the maximum PD becomes about 0.35 in the arm and 0.2 in the inter-arm.  
Although turbulent components inside the arm and inter-arm are of similar strength, mean components 
whose direction turns to the same direction in the arm region become larger than the inter-arm region. 
Therefore, frequency independent depolarization in the arm becomes smaller than that of the inter-arm and the saturation value of PD in the arm becomes larger than the value in the inter-arm region.

\subsection{Edge-on View}

\begin{figure*}
\begin{center}
\includegraphics[width=15cm]{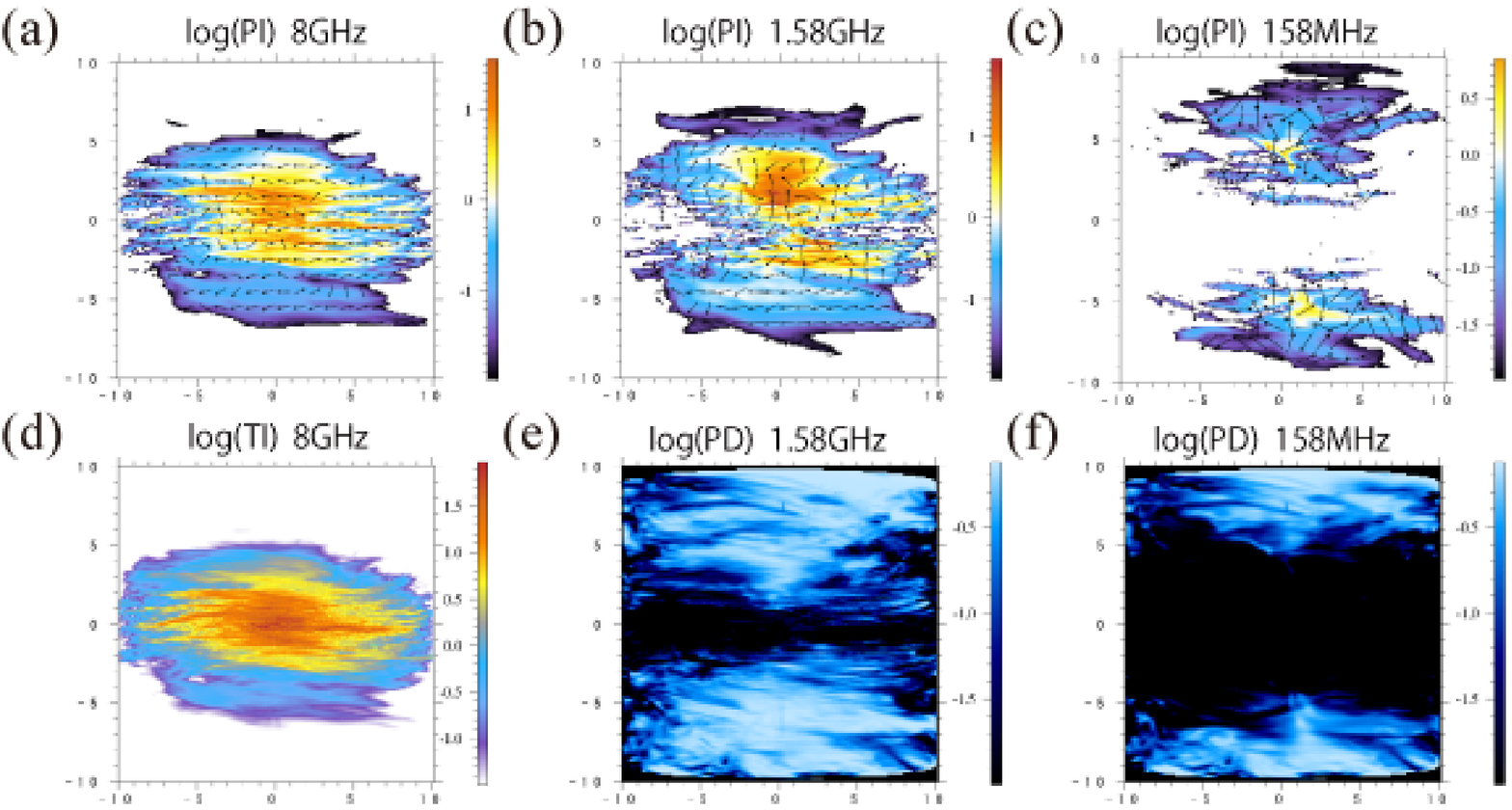}
\end{center}
\caption{
Same as Fig.\,\ref{fig3} but for $\theta=85 \degr$.
}\label{fig5}
\end{figure*}

Figs.\,\ref{fig5}a, \ref{fig5}b, \ref{fig5}c, \ref{fig5}d, \ref{fig5}e and \ref{fig5}f show the maps of PI(8\,GHz), PI(1.58\,GHz), PI(158\,MHz), TI(8\,GHz), PD(1.58\,GHz) and PD(158\,MHz), respectively.  
Overall, TI becomes larger toward the galactic mid-plane, because the global magnetic field is strongest in the disk. Structures seen in PI(8\,GHz) are similar to those seen in TI, and the magnetic vector nicely traces the global azimuthal magnetic field. Such similarity is completely missing in low frequencies. At 158\,MHz, PI is seen mostly in the halo in spite of strong magnetic fields in the disk. 
The direction of magnetic vectors becomes vertical as a result of Faraday rotation. The shape of PI(1.58\,GHz) looks like an hourglass due to depolarization around the disk. The PD(158\,MHz) becomes less than $10^{-3}$ even in the halo.  
Overall, the nature of these features is the same as that seen in the face-on view.

\begin{figure}
\begin{center}
\includegraphics[width=8cm]{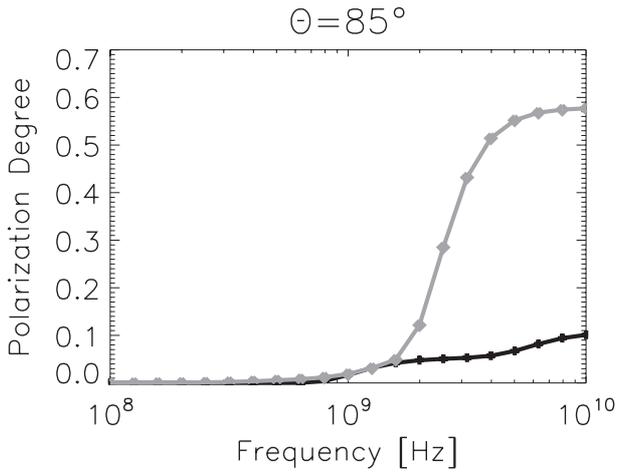}
\end{center}
\caption{
PDs of the emissions from a disk ($3$\,kpc, $0$\,kpc) (black line with crosses) and from a halo ($-3$\,kpc, $3$\,kpc) (gray line with asterisks) of $\theta=5 \degr$, as a function of frequency. 
}\label{fig6}
\end{figure}

Fig.\,\ref{fig6} shows the frequency dependence of the PD. For the LOS toward the disk (3\,kpc, 0\,kpc) for one pixel,
depolarization is significant in all frequency ranges considered; the PD at 10\,GHz becomes under 0.1 due to frequency independent depolarization. The PD falls much below 0.1 at a frequency below 1\,GHz due mostly to the external Faraday dispersion depolarization. 

Meanwhile, for the LOS toward the halo (3\,kpc, 3\,kpc), depolarization becomes important only below a few hundred MHz. This is due to the fact that the FD (and the standard deviation of the FD) is small because of the low density in the halo. 
The high saturation values of PD above 2\,GHz in the halo indicate that the mean fields become dominant, because of the outflow from the galactic disk.  

Because the global azimuthal magnetic field becomes less important at low frequencies, PI(158\,MHz) unveils secondary components of magnetic fields. For instance, the magnetic vector seems to be random, tracing not the global azimuthal magnetic field but the turbulent magnetic field. Actually, the power spectrum of magnetic fields in the halo is almost a power law at $n>2$, meaning that the halo magnetic field is turbulent. It is also interesting that, around the rotation axis of the galaxy, the magnetic vector partly traces the global vertical magnetic field and points towards the vertical direction.

\section{Discussion}

\subsection{Uncertainty of Visualization}

We have conducted observational visualization of disk galaxies using the data of numerical simulations for galactic gaseous disks. Looking forward to future broadband radio polarimetry, we have attempted to reproduce observables in a very wide range of frequencies between 100\,MHz and 10\,GHz. It is difficult to handle depolarization effects at small spatial scales, and we have adopted a model that assumes that the power of turbulent magnetic field within the spatial resolution can be replaced with that around the resolution. This adds an uncertainty to our visualization of observables at low frequencies. 

\begin{figure}
\begin{center}
\includegraphics[width=8cm]{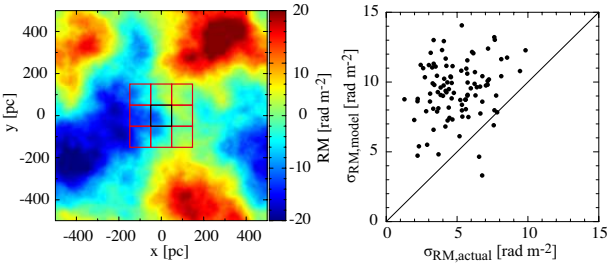} 
\end{center}
\caption{
(Left) An example of a $500^2$ pixel FD map showing Kolmogorov turbulence. The black box is used to calculate $\sigma_{\rm FD, actual}$, while $\sigma_{\rm FD, model}$ is evaluated in the red boxes (see text). (Right) Comparison between $\sigma_{\rm FD, actual}$ and $\sigma_{\rm FD, model}$. The results for 100 realization simulations are shown.
}\label{fig7}
\end{figure}

In order to estimate the uncertainty on the model, we performed independent simulations of FD maps. We made a FD map of $1$\,kpc$^2$ with $500^2$ pixels (so that the pixel resolution is 2\,pc). The FD map contains Kolmogorov turbulence with an average and standard deviation of 0\,${\rm rad \, m^{-2}}$ and 10\,${\rm rad \, m^{-2}}$, respectively. A $100$\,pc$^2$ box (the same as the resolution of our work) at the center is chosen to derive the standard deviation of FD, $\sigma_{\rm FD, actual}$. We also calculate the average FDs of eight boxes around the central box (see the left panel of Fig.\,\ref{fig7}) and derive the standard deviation of the averages, $\sigma_{\rm FD, model}$. The comparison between $\sigma_{\rm FD, actual}$ and $\sigma_{\rm FD, model}$ for 100 realizations is shown in the right panel of Fig.\,\ref{fig7}. Although there is variance between 
 the runs, the average $\sigma_{\rm FD, actual}$ and $\sigma_{\rm FD, model}$ are $\sim 5.0$ and $\sim 9.6$, respectively. Therefore, the model can estimate the standard deviation within a factor of about 2. We leave this offset in this work, since the value highly depends on the actual magnetic-field power spectrum.

Another uncertainty in our visualization is caused by numerical interpolation (averaging) around the resolution. When we obtained the average and turbulent components of FD by Eqs. (1) and (2), we adopted the average of magnetic fields inside the $\pm N+1$ cells. We have checked the dependence of the results on $N$ by varying $N=0, 1, 3, 5, 6, 10$, where $N=0$ means the interpolation with the nearest 8 cells in the cylindrical coordinate dataset. For a physical value along the LOS of the $i$-th pixel ($X_i$), the results with $N=0$ and $N=1$ are similar to each other. $N=6$ makes the power spectrum at $n>30$ steeper, and $N=10$ almost cancels out turbulent components. As for an average $X_{{\rm ave},i}$ derived with $X_j$ of eight pixels around the $i$-th pixel, $N= 3$ and $N= 6$ produce similar results to each other. Based on these studies, we decided to adopt $N=0$ in $X$ and $N=5$ in $X_{\rm ave}$.

\subsection{Dependence of the Unit Density}

\begin{figure*}
\begin{center}
\includegraphics[width=16cm]{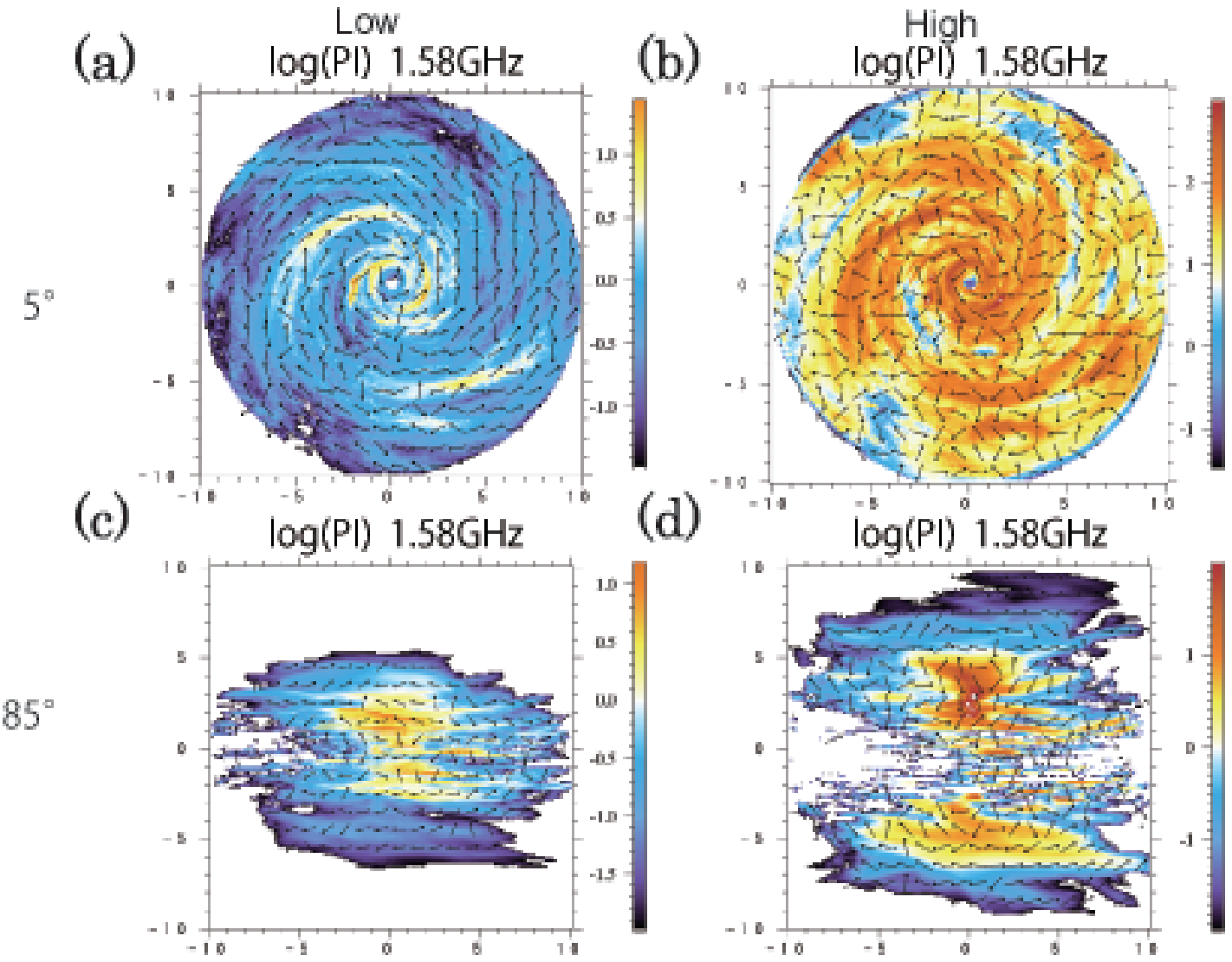}
\end{center}
\caption{
PI maps at 1.58\,GHz. Left and right panels show the results of the lower ($0.3\,\rho_0$) and higher ($3\,\rho_0$) unit densities, respectively, where $\rho_0=0.03$\,${\rm cm^{-3}}$. Top and bottom panels show the cases for $\theta = 5\degr$ and $85\degr$, respectively. 
}\label{fig8}
\end{figure*}

In the previous section, we presented the results with a fiducial unit density ($\rho_0=0.03$\,${\rm cm^{-3}}$),  
whose PDs in 100\,MHz bands are still higher than the observational results  
{\citep{mul2014, mul2018}.
It is also useful to study the cases with lower and higher unit densities, which could be compared to the spiral galaxies, respectively. Fig. \,\ref{fig8} show the results with the lower ($0.3\,\rho_0$, left panels) and higher ($3\,\rho_0$, right panels) unit densities.

In the lower case, structures of PI(1.58\,GHz) for $5\degr$ (Fig.\,\ref{fig8}a) and for $85\degr$ (Fig.\,\ref{fig8}c) are broadly consistent with those of PI(8\,GHz) with the fiducial unit density, Fig.\,\ref{fig3}b and Fig.\,\ref{fig5}b, respectively. This indicates one to one correspondence with FD (or $\sigma_{\rm FD}$) and the wavelength squared (e.g., Eqs. 6, 7, 12). In the higher case, structures of PI(1.58\,GHz) for $5\degr$ (Fig.\,\ref{fig8}b) and for $85\degr$ (Fig.\,\ref{fig8}d) are broadly consistent with those of PI(1.58\,GHz) with the fiducial unit density, Fig.\,\ref{fig3}b and Fig.\,\ref{fig5}b, respectively, except a bit stronger depolarization toward the disk plane for $85\degr$. Clear difference between the fiducial and higher unit densities is the magnetic vector; the higher case shows more random magnetic vectors due to the Faraday rotation. The high case of $85\degr$ shows the hourglass distribution and vertical magnetic vectors around the rotation axis. These structures are observed in edge-on galaxies around 1.4--4.8\,GHz. 
Because the magnetic energy in the halo region has a half of that in the disk, PI shows the characteristic feature around the rotation axis. 
Because the magnetic field strength is proportional to the number density, the strength of the PI depends on the 1.5 square of the density.  
And the PI becomes 5 times larger than the fiducial model.
Therefore, the PI in the high latitude region becomes bright in high density model.

\subsection{LOS Distribution of the dispersion of FD}

\begin{figure*}
\begin{center}
\includegraphics[width=16cm]{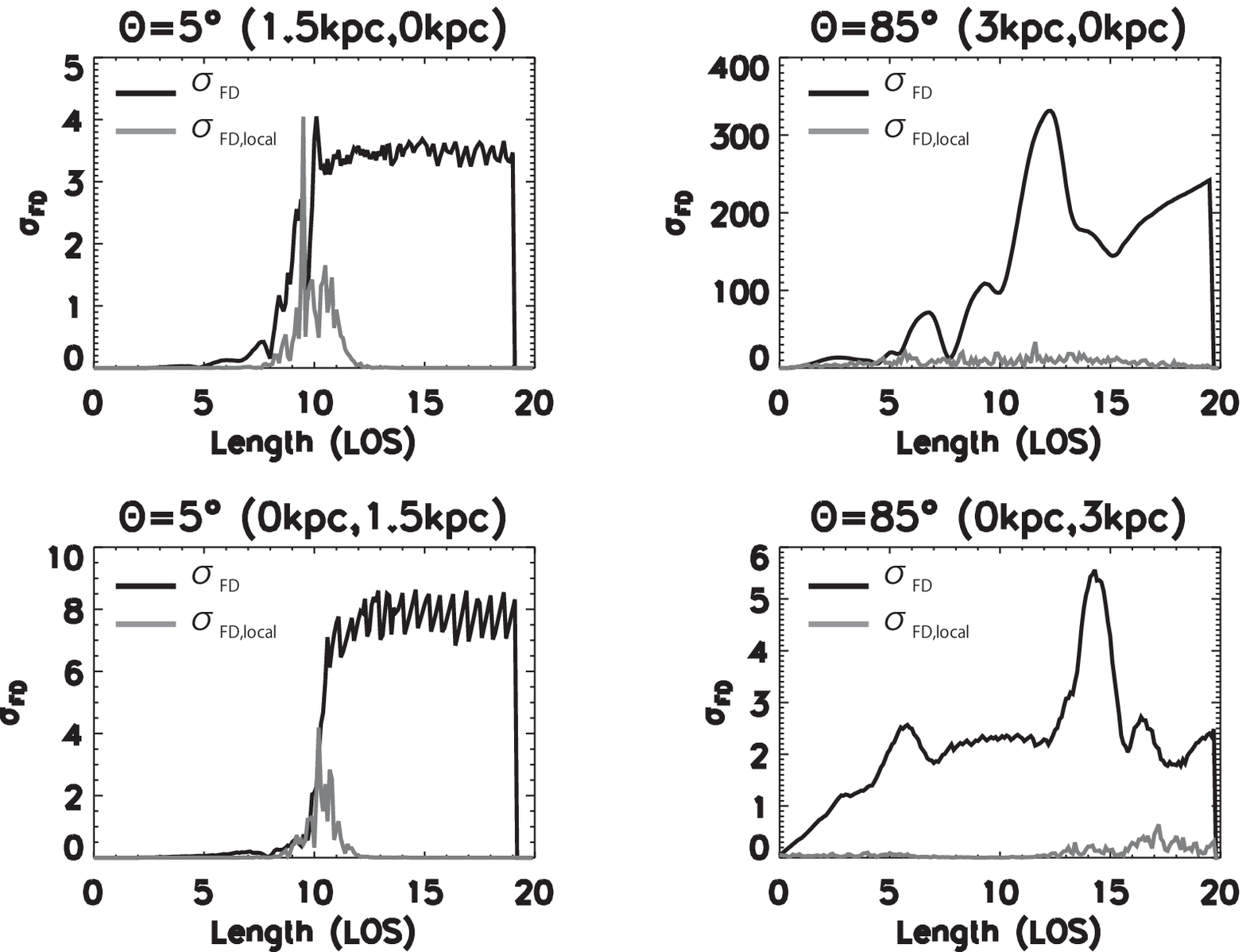} 
\end{center}
\caption{Distribution of the FD dispersion along the LOS.  Black and gray curves show the $\sigma_{{\rm FD},i,k}$ and $\sigma_{{\rm FD},i,l}$. 
 (a) top and bottom panels denote the values in the arm and inter-arm region for inclination angle $\theta = 5\degr$. 
(b) top and bottom panels show the values in disk and halo for inclination angle $\theta =85\degr$. }
\label{fig9}
\end{figure*}

Figure \ref{fig9}a shows the distribution of the FD dispersion along the LOS for the $5\degr$ inclined model.  Black and gray curves show the $\sigma_{{\rm FD}, i, k}$ and $\sigma_{{\rm FD}, i, l}$, respectively.  Top and bottom panel show the results at (2\,kpc, -0.5\,kpc), and (3\,kpc, 0\,kpc), respectively.  
The effect of the internal depolarization shows the similar tendency and the averaged values of $\sigma_{\rm FD,l}$ in the disk is about 4\,${\rm rad \, m^{-2}}$  both in the arm and the inter-arm.   On the other hand, the effect of the external depolarization is different in the arm and the inter-arm. 
$\sigma_{{\rm FD}, k}$ in the arm is comparable to the $\sigma_{{\rm FD},l}$. Because of the mean fields, FDs have similar values in the arm. Therefore, $\sigma_{{\rm FD},k}$ becomes small in the arm. The turbulent component is dominant in the inter-arm, however, FD in the inter-arm is smaller than that in the arm and can have various values.  Then $\sigma_{{\rm FD},k}$ becomes large. 
When LOS goes to the opposite side of the equatorial plane, the values of FD becomes small because of the low density. 
Therefore, the effect of the internal depolarization becomes important inside the galactic disk.  
Top and bottom panel of figure \ref{fig9}b show the results at (3\,kpc, 0\,kpc) and (3\,kpc, 3\,kpc) for the $85\degr$ inclined model, respectively, which correspond to disk and halo.  The averaged value of $\sigma_{{\rm FD}, k}$ and $\sigma_{{\rm FD},l}$ in the disk is about $200\,{\rm rad \, m^{-2}}$ and $40\,{\rm rad \, m^{-2}}$, respectively.  
When the edge-on case, the toroidal components which is the dominant components of the magnetic field in the system become the main components of the LOS fields. Of cause the turbulent components of the toroidal (azimuthal) magnetic fields  are stronger than the mean fields, the mean fields becomes 70\% of the toroidal fields. Therefore, the effect of the internal depolarization becomes a smaller contribution than that of the face-on case.

\subsection{Can PI at MHz-band be observed or not?}

\begin{figure*}
\begin{center}
\includegraphics[width=16cm]{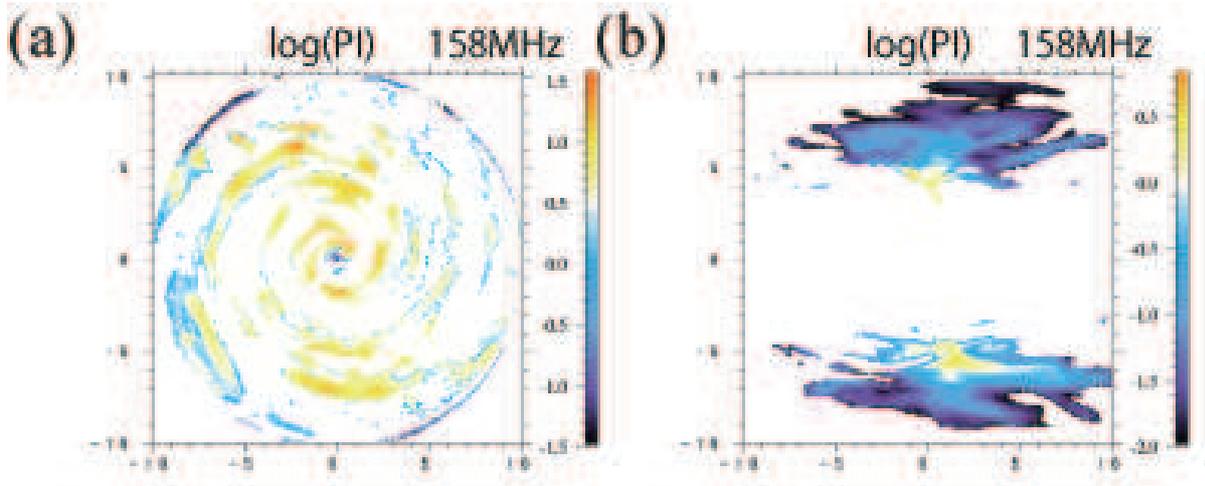} 
\end{center}
\caption{PI maps masked where PD is over 0.01. (a) $\theta =5\degr$, (b) $\theta=85\degr$.
 }
\label{fig10}
\end{figure*}

Fig.\,\ref{fig10} shows the PI at the observing frequency of 158\,MHz masked where PD\,$< 0.01$.  Because Faraday depolarization works effectively in magnetic spiral arms, PD in the arm becomes lower than 0.01 in the case of $5\degr$.  
The remaining PI components are created in the halo and maximum strength of PI is 38  $\mu {\rm Jy/beam}$ (Fig. \ref{fig10}a).  On the other hand, in the case of $85\degr$, PI are remained around the rotational axis and its strength is less than $10 \mu {\rm Jy/beam}$.

\section{Summary}

We summarize the points discussed in the present work. 

\begin{itemize}
\item Total synchrotron intensity (TI) and PI(>4\,GHz) are both stronger along denser magnetic spiral arms, where depolarization is insignificant at such high frequencies. The polarization angle (PA) traces that of the azimuthal magnetic field inside the disk. Meanwhile, PI(<1\,GHz) through the disk weakens due to depolarization. Note that the maximum values of PI(>158\,MHz) and PI(>8\,GHz) are similar to each other, though they appar at different places.  
An increase of PI from 8\,GHz to 1\,GHz are shown in our pseudo observations in the whole disk, althugh such increase was observed only in the our regions of 
some face-on galaxies. This result may sugget that the depolarization effect in our model is underestimated for the real galaxies. 

\item PI(158\,MHz) through the magnetic spiral arms is faint. This is because depolarization significantly occurs due to large FD. Consequently, PI at such low frequencies traces the magnetic field structure in the halo. These tendencies do not depend on the inclination angle of the galaxy. The high PI(158\,MHz) in the edge-on view traces the halo magnetic field formed by the intermittent emerging of magnetic flux produced by the Parker instability.

\item When the inclination angle is $5\degr$, because the local FD dispersion inside the disk has a similar value to the FD dispersion, the effect of the internal Faraday depolarization becomes important inside the disk. On the other hand, in the case of $85\degr$, FD dispersion becomes over 10 times larger than the local FD dispersion.  
\end{itemize}

In this paper, we did not employ Faraday tomography, which is clearly complementary to our work and may be more powerful to deproject the LOS structure. It is, however, not trivial so far how we can translate the Faraday spectrum into real 3D structure of the galaxy \citep{ide2014}. We need further studies of both observational visualization and Faraday tomography for numerical data.

\section*{Acknowledgements}
We are grateful to Dr. R. Matsumoto, Dr. K. Takahashi, and Dr. S. Ideguchi for useful discussion. 
We also thank the anomolous referee for his/her userful comments and costructive suggestions. 
Numerical computations were carried out on SX-9 and XC30 at the Center for Computational Astrophysics, CfCA of NAOJ (P.I. MM). A part of this research used computational resources of the HPCI system provided by (FX10 of Kyushu University) through the HPCI System Research Project (Project ID:hp140170).This work is financially supported in part by a Grant-in-Aid for  Scientific Research (KAKENHI) from JSPS (P.I. MM:23740153, 16H03954,  TA:15K17614, 15H03639).


\bsp	
\label{lastpage}
\end{document}